# Hybrid 2D-QD MoS$_2$-PbSe Quantum Dot Broadband Photodetectors with High-Sensitivity and Room-Temperature Operation at 2.5 µm

*Biswajit Kundu$^\perp$, Onur Özdemir$^\perp$, Mariona Dalmases, Gaurav Kumar and Gerasimos Konstantatos\**

Dr. B. Kundu, O. Özdemir, Dr. M. Dalmases, G. Kumar
ICFO - Institut de Ciències Fotòniques, Av. Carl Friedrich Gauss 3, 08860 Castelldefels (Barcelona), Spain

Prof. G. Konstantatos
ICFO - Institut de Ciències Fotòniques, Av. Carl Friedrich Gauss 3, 08860 Castelldefels (Barcelona), Spain
ICREA - Institució Catalana de Recerca i Estudis Avançats, Passeig Lluís Companys 23, 08010 Barcelona, Spain

E-mail: gerasimos.konstantatos@icfo.eu

Broadband infrared photodetectors have profound importance in diverse applications including security, gas sensing, bioimaging, spectroscopy for food quality, and recycling, just to name a few. Yet, these applications can currently be served by expensive epitaxially grown photodetectors, limiting their market potential and social impact. The use of colloidal quantum dots (CQDs) and 2D-materials in a hybrid layout is an attractive alternative to design low-cost CMOS-compatible infrared photodetectors. However, the spectral sensitivity of these conventional hybrid detectors has been restricted to 2.1 µm. Herein, we present a hybrid structure comprising MoS$_2$ with PbSe CQDs to extend their sensitivity further towards the mid-wave infrared, up to 3 µm. We achieve room temperature responsivity of 137.6 A/W and a detectivity of $7.7 \times 10^{10}$ Jones at 2.55 µm owing to highly efficient photoexcited carrier separation at the interface of MoS$_2$ and PbSe in combination with an oxide-coating to reduce dark current; the highest value yet for a PbSe based hybrid device. These findings strongly support the successful fabrication of hybrid devices which may pave the pathway for cost-effective, high performance, next-generation, novel photodetectors.

# 1. Introduction

Low-cost, high sensitivity infrared photodetectors are vital components in various applications such as material sorting, telecommunications, and bioimaging and cover a spectral range from short-wave infrared (SWIR, 1-2 μm) to mid-wave infrared (MWIR, 2-5 μm) and long-wave infrared (LWIR, 5-12 μm).[1–4] Currently, the field of infrared photodetectors is dominated by mature material technologies such as epitaxially grown III-V materials, ternary systems like HgCdTe (MCT), quantum wells, and type-II superlattices structures.[1,5,6] However, the fabrication of these photodetectors is relatively complex and expensive with a low yield.[7] In many applications, they also require cooling at cryogenic temperatures to minimize the thermal background and to obtain low noise and high sensitivity, typically in MWIR and LWIR wavelength regimes.[1] Therefore, their use is mostly limited to high-end military and scientific applications, while a huge possibility exists for their widespread use in several civil consumer applications as well as in biomedicine, industrial inspection, food quality control, and recycling.

Colloidal quantum dots (CQDs) are an attractive low-cost alternative to conventional epitaxial semiconductors.[8] Precise and simple control of their bandgap,[9–12] as well as ease of tuning their electrical and optical properties by modifying their surface chemistry provides a versatile active material for photodetectors. Easy processibility and compatibility with the well-developed silicon technologies[13] makes CQDs a low-cost alternative for numerous applications such as photodetectors[14], light-emitting diodes[15], lasers[16], solar cells[17] and field-effect transistors[18].

Lead chalcogenide CQDs, like PbS and PbSe, have been widely explored in photodetector applications owing to their wide range of bandgap tunability from the visible to infrared.[11,19] Still, low carrier mobilities in these QD films ($10^{-3}$-1 cm$^2$ V$^{-1}$ s$^{-1}$)[20,21] impose constraints on the magnitude of photoconductive gain in photoconductive detectors. To overcome this mobility-bottleneck, hybrid devices combining CQDs with high mobility 2-dimensional (2D) materials, such as transition metal dichalcogenides (TMDCs) and graphene, come into play.[22] 2D materials have been proven to enhance performance in many hybrid photodetector systems.[23–28] High mobility graphene (with mobilities around $10^3$-$10^4$ cm$^2$ V$^{-1}$ s$^{-1}$) and TMDCs ($10^1$-$10^3$ cm$^2$ V$^{-1}$ s$^{-1}$) were proven useful in the efficient transfer of photogenerated charges that led to the development of fast and sensitive hybrid detectors.[29,30] The first high sensitivity ($10^7$ A/W) hybrid photodetector operating around 950 nm wavelength is based on PbS QDs and graphene where graphene acts as a high mobility carrier transport channel for the charges generated by PbS QDs.[31]

As an alternative to graphene, TMDCs offer a unique opportunity to overcome the dark current challenge that rises from the semi-metallic nature of graphene. In contrast to zero bandgap graphene, high mobility semiconducting TMDCs exhibit a tunable bandgap in the range of 1-2 eV, making it a suitable candidate for high sensitive photodetection with low off-state currents.[32] The use of semiconducting 2D TMDC channel in hybrid devices is of particular interest for setting up an appropriate band offset with the energy levels of QDs, which depend not only on the bandgap but also on the work function of the individual component. Among the TMDCs, molybdenum disulfide (MoS$_2$) is a prominent candidate as a high-mobility transistor channel material. The in-plane mobility of single-layer MoS$_2$ has been reported as high as 200 cm$^2$ V$^{-1}$ s$^{-1}$ after encapsulation with HfO$_2$ and exceeded the current on/off ratios of $10^8$ at room temperature.[33] The combination of MoS$_2$ with CQDs has been demonstrated to exhibit photoconductive gains of $10^5$-$10^7$, a consequence of an internal gain mechanism with holes staying in the QDs while electrons move multiple times in the circuitry under illumination. The resulting detectivities are on the order of $10^{12}$ Jones in the spectral range from 700 nm to 1200 nm.[34] A recent MoS$_2$-QDs hybrid device has demonstrated high sensitivity up to 2.1 µm.[35] However, accessing the spectral sensitivity above 2 µm from a QD-based hybrid device remains a challenge due to the limited choice of available sensitizers in this spectral region.

In this paper, we demonstrate a MoS$_2$-PbSe QD hybrid photodetector that has a broadband response up to 3 µm at room temperature. The efficient charge separation at the MoS$_2$-PbSe interface, the high mobility of the MoS$_2$ along with oxide-isolation of the contacts, allowed us to reach a responsivity of 137.6 A/W and a detectivity of $7.7 \times 10^{10}$ Jones at 2.55 µm, which is the highest reported to date for QD detectors at those wavelengths.

## 2. Results and Discussion

### 2.1. PbSe Quantum Dots

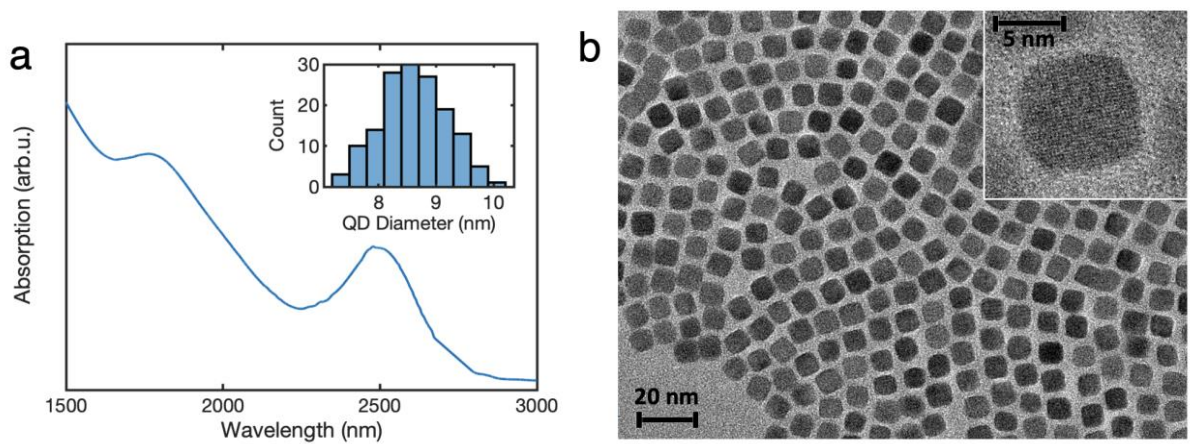

**Figure 1.** (a) Absorption versus wavelength of the PbSe CQD in tetrachloroethylene solution, with a prominent peak around 2.5 µm. Inset: Size distribution of the QDs obtained from TEM measurements. The average diameter of the QDs is 8.6 ± 0.57 nm. (b) TEM image of the PbSe QDs after synthesis. Scale bar is 20 nm. Inset: A magnified TEM image of a single PbSe QD with a diameter around 9 nm.

**Figure 1**a shows the optical absorption spectrum of oleic acid capped PbSe CQDs dispersed in tetrachloroethylene solution. A clear excitonic peak at 2.5 µm (0.5 eV) along with a secondary peak around 1.8 µm (0.69 eV) is visible. A transmission electron microscope (TEM) image of the synthesized QDs is presented in Figure 1b, which shows the non-agglomerated and uniform distribution of QDs. Inset in Figure 1b reveals the single

crystalline structure of a single PbSe QD with a higher magnification. The average diameter of these QDs is measured to be 8.6 nm with a standard deviation of 0.57 nm. The QDs have radii smaller than the excitonic Bohr radius of 46 nm,[36] which indicates that the QDs we have synthesized are in the quantum confinement regime and further confirms the excitonic peaks in the absorption spectra.

## 2.1. PbSe QD based Photodetector

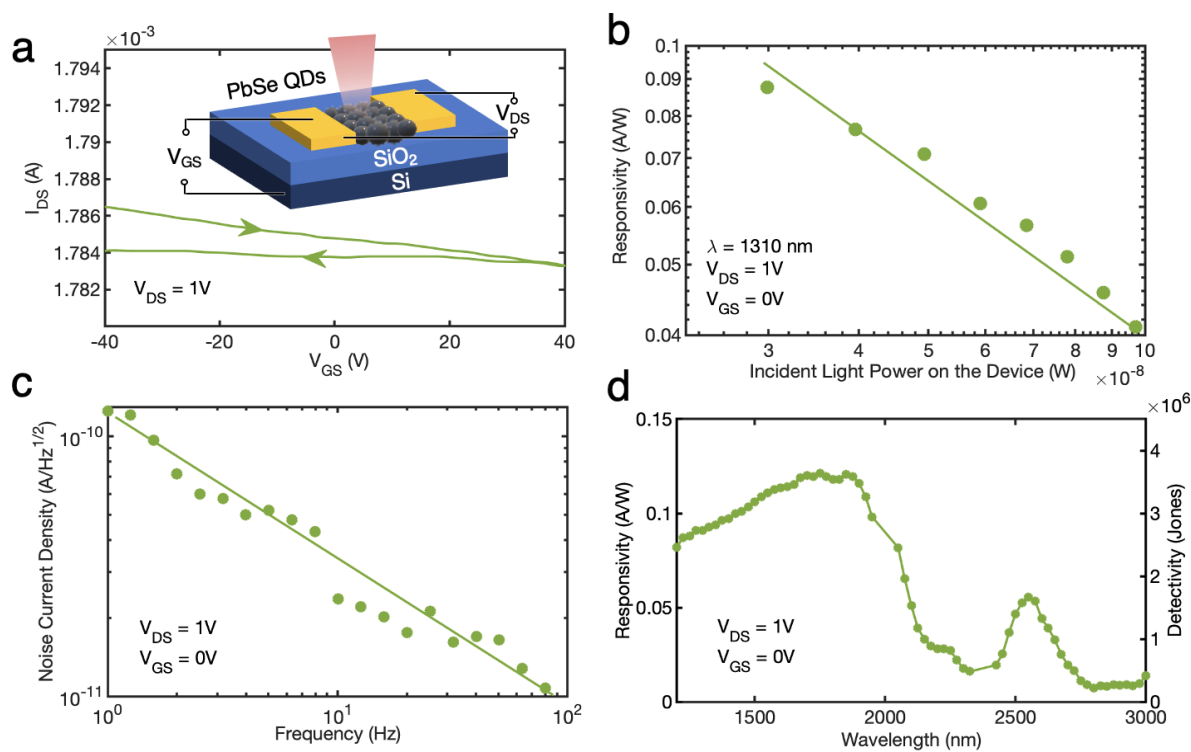

**Figure 2.** (a) $I_{DS}$ versus $V_{GS}$ curve for the PbSe photodetector without a MoS$_2$ layer below with $V_{DS} = 1V$. Arrows indicate the direction of the $V_{GS}$ sweep. A 3D schematic of the device can be seen as an inset. (b) Responsivity of the device with varying light intensity. The device is illuminated with a 1310 nm laser and with $V_{DS}$ of 1V. A solid green line is a guide to the eye and shows a logarithmic dependence. (c) Noise spectral density in dark with $V_{DS} = 1V$. Solid green line is a guide to the eye, depicting the 1/f behavior. (d) Spectral responsivity and detectivity (D*) of the PbSe photodetector between the wavelengths of 1.2 - 3 μm.

In order to utilize PbSe QDs as functional active materials in photodetectors, we first characterized the PbSe photodetector without a TMDC layer. Inset in **Figure 2**a shows a device schematic with a back-gated device architecture. We spincoated PbSe QDs on top of a

doped-Si/SiO$_2$ substrate with gold source-drain electrodes. In order to form a close-packed film with high mobility, we used 1,2-ethanedithiol (EDT) to replace the long oleic acid ligands on the surface of the QDs during spin coating. Source-drain current (I$_{DS}$) versus backgate voltage (V$_{GS}$) curve can be seen in Figure 2a. The PbSe film shows a p-type behavior with a field-effect hole mobility of 0.96 cm$^2$ V$^{-1}$ s$^{-1}$.

Figure 2b shows the light-intensity dependent responsivity of the PbSe photodetector to a 1310 nm laser with adjustable light intensity and exhibits a maximum responsivity of 0.08 A/W. We observe a decrease in responsivity as the incident light intensity increases as a consequence of the saturation of sensitizing traps in PbSe QDs.[37] The spectral photoresponse between 1.2 µm and 3 µm can be seen in Figure 2d. A clear peak in the responsivity is present near the excitonic absorption of the QDs, with values reaching 0.055 A/W at 2.55 µm. To characterize the noise in our device, we also performed noise current spectral density measurements with a lock-in amplifier, using the same biasing conditions as the responsivity measurements. The noise current spectral density is shown in Figure 2c with the typical 1/f component is indicated by a solid line. To put our photodetector in perspective, we calculated normalized detectivity, D*, as:

$$D^* = \frac{R\sqrt{A}}{i_n}$$

where $R$ is the responsivity in A/W, $A$ is the device area in cm$^2$, and $i_n$ is the bandwidth-normalized noise current with units of A Hz$^{-1/2}$. The PbSe QD photodetector has a detectivity of 1.96 x 10$^6$ Jones at 2.55 µm at room temperature.

**2.2. MoS$_2$ - PbSe QD Hybrid Photodetectors**

As a next step in enhancing the performance of the QD-based photodetector, we employed a hybrid structure with few-layer MoS$_2$ and PbSe QDs to take advantage of the built-in electric field and inherent gain mechanism present in similar hybrid systems.[38,39] A schematic of the hybrid device can be seen in **Figure 3**a. Briefly, we fabricated the device by transferring MoS$_2$ flakes on a doped-Si/SiO$_2$ (285 nm) substrate. Then, gold source and drain contacts were patterned on each side of the MoS$_2$ flake, creating a MoS$_2$ transistor. The thickness of the MoS$_2$ flake was confirmed by atomic force microscopy (AFM) measurements (Figure S1a-b) and match the thickness of a 6-layer MoS$_2$ flake. We utilized few-layer (5-6 layers) MoS$_2$ as it exhibits lower noise than single-layer or bulk MoS$_2$.[40] Finally, PbSe QDs were spin-coated on top of the MoS$_2$ transistor following a conventional layer-by-layer approach using EDT in the ligand exchange process. The thickness of the PbSe film is 53 nm (Figure S1c-d).

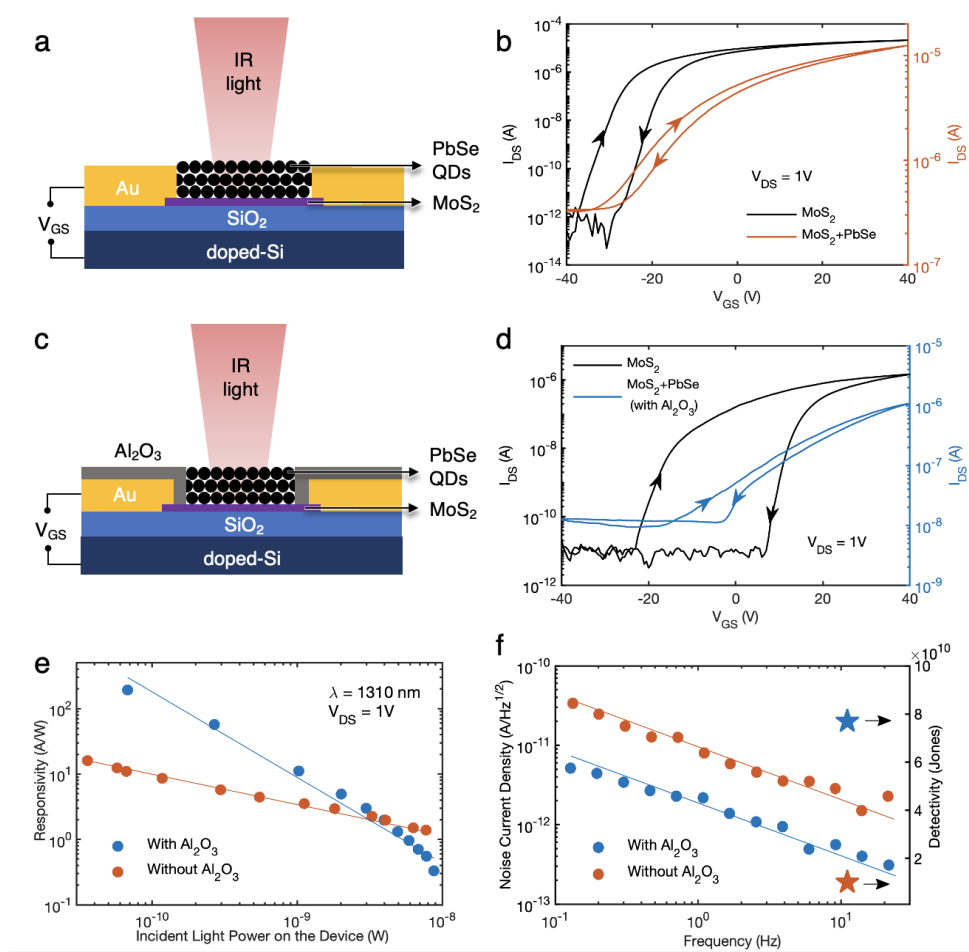

**Figure 3.** Schematic of the hybrid MoS$_2$-PbSe photodetector (a) without Al$_2$O$_3$ isolated contacts and (c) with Al$_2$O$_3$ isolated contacts under IR illumination. (b) I$_{DS}$ versus V$_{GS}$ curves of the hybrid device before (black curve, left y-axis) and after (orange curve, right y-axis) the PbSe QD deposition. Arrows indicate the direction of the V$_{GS}$ sweep. The measurements are performed in dark with $V_{DS}$ of 1V. (d) $I_{DS}$ versus $V_{GS}$ curves for the device with $Al_2O_3$ layer before (black curve, left y-axis) and after (blue curve, right y-axis) PbSe QD deposition. (e) Responsivity versus light intensity for the hybrid devices, illuminated with a 1310 nm laser. Solid lines are guides to the eye. (f) Noise spectral density in the dark for both hybrid detectors. Detectivity values (D*) are indicated by stars (blue for the oxide-covered device and orange for the device without, y-axis on the right) Solid line is a guide to the eye and shows the 1/f behavior.

Figure 3b illustrates the $I_{DS}$-$V_{GS}$ characteristics of the hybrid device before and after the QD deposition in dark. A typical transistor behavior is observed with off-state currents of around 320 nA in the final device under 1V bias. The resulting hybrid device has a field-effect electron mobility of 16.5 cm$^2$ V$^{-1}$ s$^{-1}$.

We further improved the performance of the hybrid device by coating the gold contacts with aluminum oxide ($Al_2O_3$) layer. By doing so, we blocked direct charge transfer from the QD layer to the contacts. Instead, the photogenerated carriers in the QD layer go through $MoS_2$ to reach the contacts under illumination. A schematic of the final device can be seen in Figure 3c as well as schematics for the fabrication process in Figure S3 and a description of the process in the experimental section.

Figure 3d shows the $I_{DS}$-$V_{GS}$ curves before and after the QD layer deposition in the dark with off-state currents around 10 nA, demonstrating the effect of reduced charge transfer from the QDs. Field-effect electron mobility in the final device is calculated to be 6.37 cm$^2$ V$^{-1}$ s$^{-1}$. Deposition of PbSe QDs on top of $MoS_2$ forms a *p-n* junction at the interface, as demonstrated by the ultraviolet photoelectron spectroscopy (UPS) measurements and the resulting band diagram (Figure S2). Under illumination, QD film absorbs light creating electron–hole pairs that are separated by the built-in field at the *p-n* heterojunction. Holes are trapped in the QD layer while electrons are transported by the $MoS_2$ channel to the contacts, leading to the aforementioned inherent gain mechanism.

Figure 3e shows the light intensity dependent responsivity of the hybrid devices to a 1.3 µm laser. For the hybrid devices with and without an oxide layer, the responsivities reach up to 192 A/W and 23.5 A/W respectively. At higher light intensities, a large number of photo-generated carriers induce a reverse electric field, effectively decreasing the built-in potential at the MoS$_2$-PbSe interface and resulting in a drop of responsivity, as seen in both devices.[41] This effect is enhanced for the device with the oxide layer, as the photocurrent leakage from the QD layer to the contacts is reduced. This ensures that most of the current pass through the MoS$_2$-PbSe interface, leading to a higher responsivity at low light intensities but also a faster build-up of a reverse electric field with increasing light intensity.

Noise spectral density measurements for both hybrid detectors in dark at room temperature are shown in Figure 3f with the typical 1/f behavior indicated by solid lines. Al$_2$O$_3$ covered hybrid device exhibits lower noise across the frequencies of measurement. This results in a detectivity (D*) of 1.0 x 10$^{10}$ Jones for the device without Al$_2$O$_3$ and 7.7 x 10$^{10}$ Jones for the device with Al$_2$O$_3$.

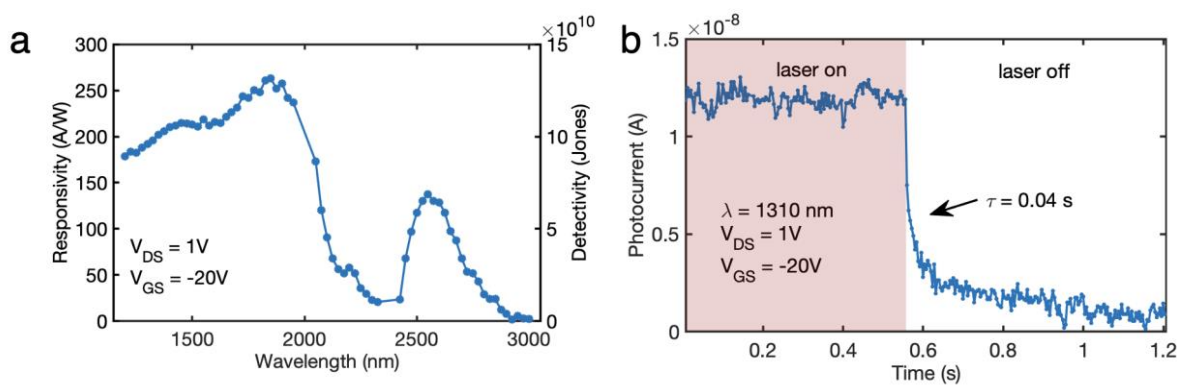

**Figure 4.** Spectral responsivity and detectivity (D*) of the hybrid photodetector. A clear peak around 2.5 µm shows the response from the PbSe QDs. (b) Response speed of the hybrid photodetector. When the 1310 nm laser illumination is turned off, photocurrent diminishes exponentially with a time constant of 0.04s.

We further characterized the device with Al$_2$O$_3$ by measuring the spectral responsivity in a wavelength range from 1.2 µm to 3 µm as shown in **Figure 4**a. With a prominent excitonic peak, the device has a responsivity of 137.6 A/W at 2.55 µm, which leads to the detectivity of 7.7 x 10$^{10}$ Jones.

Figure 4b shows the time response of our device as the laser illumination is turned off to determine how fast our device can respond to incident light. Photocurrent shows an exponential decay with a time constant, $\tau_{lifetime}$, of 0.04 seconds. Internal photoconductive gain, $G$, which is an outcome of photogenerated electrons circulating many times in the device while holes are trapped in the QDs [42], can then be calculated as $G = \frac{\tau_{lifetime}}{\tau_{transit}}$. $\tau_{transit}$ is the carrier transfer time through the contact and can be calculated as $\tau_{transit} = \frac{L^2}{\mu V_{DS}}$, where $L$ is the channel length, $\mu$ is the FET mobility, and $V_{DS}$ is the source-drain voltage. For our hybrid device with oxide, the photoconductive gain is in the order of $10^6$.

## 3. Conclusion

In conclusion, we demonstrated high sensitivity, broadband infrared hybrid $MoS_2$-PbSe photodetector that operates at room temperature with a responsivity of 137.6 A/W and a detectivity of $7.7 \times 10^{10}$ Jones at 2.55 µm. Dark current reduction strategies like $Al_2O_3$ coating the electrical contacts in combination with the built-in electric field and an inherent gain mechanism allowed us to have a photodetector, highest reported to date for QD detectors at those wavelengths, with responses up to 3 µm. These thin hybrid devices have the potential to be useful in low-cost high-performance applications without any cooling and can be an alternative to current technologies.

## 4. Experimental Section

*Chemicals:* Lead oxide (PbO, 99.999%), oleic acid (OA, tech. grade 90%), diphenylphosphine (DPP, 98%), butylamine (99.5%), and 1,2-ethanedithiol (EDT) were purchased from Sigma Aldrich. 1-octadecene (ODE, tech. grade 90%) was purchased from Alfa Aesar. Selenium powder (Se, 99.99%) and trioctylphophine (TOP, 97%) were purchased from Strem Chemicals. All reagents and solvents were used as received.

*Synthesis of the PbSe Colloidal Quantum Dots:* 2.5 µm excitonic peak based PbSe QDs were synthesized under inert atmosphere by hot injection method. 2 mmol PbO, 15 mmol OA, and 10 mL ODE were mixed in a three-neck flask and heated at 100 °C, under vacuum, for 1 hour to form lead oleate, stirring at 650 rpm. Thereafter, the solution was placed under Ar, the temperature was raised to 180 °C and a solution of TOP-Se was quickly injected (2 mmol Se, 200 µL DPP and 1 mL TOP). After 3 minutes of reaction under these conditions, a second solution of TOP-Se (2 mmol Se in 2 mL TOP) was injected dropwise. Immediately after this second injection, the reaction was quenched with a water bath. Once at room temperature, QDs were separated by centrifugation and redispersed in anhydrous toluene. The addition of 10µL butylamine was necessary to redisperse the precipitate. Subsequently, the product was washed for three times with a mixture of acetone/ethanol and finally redispersed in anhydrous toluene, adjusting the concentration to 30 mg/mL. The final solution was bubbled with $N_2$ to avoid the oxidation of the material.

*Optical Absorption and TEM Measurements:* Absorption measurements were made under ambient conditions, using a Cary 5000 UV-Vis-NIR. TEM images were acquired in a JEOL JEM 2100 equipped with a $LaB_6$ source.

*MoS$_2$ Field-Effect Transistor (FET) Fabrication:* Commercially available bulk MoS$_2$ crystal was purchased from 2D semiconductor cooperation. The Si/SiO$_2$ (285 nm) substrates were first cleaned in acetone and isopropanol in an ultrasonic bath. Then, MoS$_2$ flakes were exfoliated onto the Si/SiO$_2$ substrate using a polydimethylsiloxane (PDMS) tape. After the transfer, the substrate was washed with acetone and isopropanol to remove possible residue from the PDMS tape. Regions of few-layer MoS$_2$ flakes were identified under the optical microscope. Using the AZ5214E photoresist, source and drain contacts were patterned with photolithography (Maskless Aligner Heidelberg MLA 150). After development, these regions are coated with Ti (3 nm)/Au (50 nm) with e-beam and thermal deposition. After the lift-off to reveal the metallic contacts, the samples were annealed under an inert atmosphere at 150 °C for 3 hours to improve the contact adhesion to MoS$_2$. Second photolithography was performed for the oxide-isolated devices. A 3 µm by 12 µm rectangular area on top of the MoS$_2$ flakes as well as a bigger 500 µm x 500 µm area on the contact pads far away from the active device area is patterned using Maskless Aligner Heidelberg MLA 150 with the ECI3007 photoresist to act as a protective layer for the oxide deposition. After the development of the exposed regions, 5 nm of Al$_2$O$_3$ is coated onto the device with atomic layer deposition (Arradiance GEMStar). Liftoff of the photoresist is performed in hot (60 °C) acetone overnight followed by isopropanol wash.

*PbSe QDs Film Deposition and Ligand Exchange:* PbSe QDs film was prepared by the conventional layer-by-layer technique using the solid-state ligand exchange procedure at room temperature. A layer of oleic acid capped PbSe CQDs in toluene (30 mg/mL) was first spin-coated at 3000 rpm. Then 6 drops of 1,2-ethanedithiol (EDT) solution (0.2% in acetonitrile) were dropped onto the substrate. After 25 s the sample is again let to spin at 2500 rpm for 10 s. After that, to remove the excess unbound ligands, films were washed followed by three rinsing steps with acetonitrile. The whole process was repeated two times to achieve a thickness of the film in the range of 50-60 nm.

*FET Mobility Calculations:* The field-effect mobilities, µ, of our devices are calculated under the short-channel approximation by:

$$\mu = \frac{dI_{DS}}{dV_{GS}} \frac{L}{W C_{ox} V_{DS}}$$

where $V_{DS}$ is the source-drain voltage, $I_{DS}$ is the source-drain current, $V_{GS}$ is the gate backgate voltage, L and W are the channel length and channel width, and $C_{ox}$ is the oxide capacitance of the $SiO_2$ layer ($1.21 \times 10^{-8}$ F cm$^{-2}$). In calculating the µ values, the forward transfer curves are taken (as the backgate voltage increases with respect to the source contact).

*Electrical and Optical Characterization:* All the measurements were done at room temperature under ambient conditions. $I_{DS}$-$V_{GS}$ curves of the FET devices were measured in an electromagnetically isolated probe station with a semiconductor analyzer (Keysight B1500A) using a four-probe connection. Power dependent responsivity measurements were performed in the same setup using a fiber-coupled 4-channel Thorlabs MCLS1 laser and its intensity was controlled by an Agilent A33220A waveform generator. The spectral responsivity measurements are performed in a Bentham monochromator setup with a QTH source and mechanical chopper (11 Hz) The device is contacted with 2 probes and the bias is applied through the contacts. The extracted current is amplified by a Stanford Research low-noise transimpedance amplifier (SR570) and fed into a Zurich Instruments MLFI lock-in for readout. Spectral noise current density measurements were carried out in an electrically shielded probe station system using a lock-in based setup (Zurich instruments MFLI). An external low-noise filter along with a two-channel low noise Keysight B2926A power source was used to apply the bias to the sample, and the signal to the lock-in amplifier was fed through a low-noise current amplifier (SR570).

**Supporting Information**

Supporting Information is available from the Wiley Online Library.

**Acknowledgements**

[⊥]Biswajit Kundu and Onur Özdemir contributed equally to this work.

The authors acknowledge financial support from the European Research Council (ERC) under the European Union's Horizon 2020 research (grant agreement no. 725165) as well as Graphene Flagship under Grant Agreement Nr. 881603 (Core3). The authors also acknowledge financial support from the Spanish Ministry of Economy and Competitiveness (MINECO), and the "Fondo Europeo de Desarrollo Regional" (FEDER) through grant TEC2017-88655-R, from the Spanish State Research Agency through the "Severo Ochoa" program for Centers of Excellence in R&D (CEX2019-000910-S), from Fundació Cellex, Fundació Mir-Puig, and from Generalitat de Catalunya through the CERCA program.

Received: ((will be filled in by the editorial staff))
Revised: ((will be filled in by the editorial staff))
Published online: ((will be filled in by the editorial staff))

**ToC (Table of Contents)**

We demonstrate infrared photodetectors with MoS$_2$ and PbSe quantum dot hybrids with high sensitivity at room temperature. With detectivities above 7 x 10$^{10}$ Jones at 2.5 µm, highest reported to date, these detectors are cost-effective and can be integrated with current silicon-based architectures to pave the way for future infrared technologies.

Biswajit Kundu[⊥], Onur Özdemir[⊥], Mariona Dalmases, Gaurav Kumar and Gerasimos Konstantatos*

**Hybrid 2D-QD MoS$_2$-PbSe Quantum Dot Broadband Photodetectors with High-Sensitivity and Room-Temperature Operation at 2.5 µm**

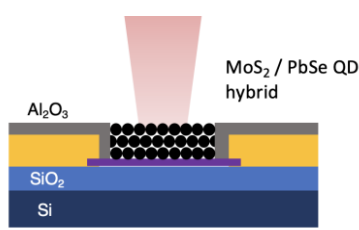